\documentstyle[12pt]{article}

\font\tenrm=cmr10
\font\tenit=cmti10
\font\tenbf=cmbx10
\font\elevenbf=cmbx10 scaled\magstep 1
\font\elevenrm=cmr10 scaled\magstep 1
\font\elevenit=cmti10 scaled\magstep 1
\font\ninebf=cmbx9

\font\nineit=cmti9

\begin{document}
% new macro for bibliography
\newcommand{\bibit}{\nineit}
\newcommand{\bibbf}{\ninebf}
\renewenvironment{thebibliography}[1]
{   \begin{list}{\arabic{enumi}.}
    {\usecounter{enumi} \setlength{\parsep}{0pt}
     \setlength{\itemsep}{2pt} \settowidth{\labelwidth}{#1.}
     \sloppy
    }}{\end{list}}

\parindent=3pc
\newcommand{\sbar}{\,\overline{\! S}}
\newcommand{\barz}{\,\overline{\! Z}}
\newcommand{\zbar}{\bar{z}}
\newcommand{\tbar}{\overline{T}}
\newcommand{\z}{\zeta}
\newcommand{\zb}{\bar{\zeta}}
\newcommand{\psibar}{\overline{\Psi}}
\newcommand{\cm}{{\elevenit Commun. Math. Phys.} }
\newcommand{\pr}{{\elevenit Phys. Rev.} }
\newcommand{\pl}{{\elevenit Phys. Lett.} }
\newcommand{\np}{{\elevenit Nucl. Phys.} }

%\begin{titlepage}
%\begin{flushright}
%\hfill{CPTH-C218.0193}\\[1mm]
%NUB--????\\[1mm]
%January 1993
%\end{flushright}
%\vskip 1cm

\begin{center}{
{\tenbf SCHWINGER EFFECT IN STRING THEORY
\footnote{Proceedings of the Rome Workshop on
{\it Quantum Gravity, String Theory and Grand Unification}, Rome (September
1992). } }
\vglue 1.0cm
{\tenrm C.BACHAS\\}
\baselineskip=13pt
{\tenit Centre de Physique Th\'eorique, Ecole Polytechnique\\}
\baselineskip=12pt
{\tenit  91128 Palaiseau, France\\}
\vglue 0.8cm
{\tenrm ABSTRACT}}
\end{center}
\vglue 0.3cm
{\rightskip=3pc
 \leftskip=3pc
 \tenrm\baselineskip=12pt
 \noindent
I discuss the quantum instability of an electric field  in a theory
of open strings.
     \vglue 0.6cm}
%\end{titlepage}
\baselineskip=14pt
%\vglue 0.4cm

\elevenrm

\vskip1cm

One of the few available hints on the interplay between
quantum theory
and gravity  is the Hawking emission of a black hole
 $^1$.
This can be described as pair creation in the background gravitational
field. Of the two produced particles one stays trapped inside
the horizon, while
the other escapes to infinity where it emerges with a thermal
spectrum at an
effective temperature $T_H = (8\pi GM)^{-1}$.
In deriving his famous result Hawking   assumed a
Schwarzild form for the metric, and minimally-coupled
non-self-interacting
quantum fields.
In string theory we expect two kinds of corrections to this result:
  {\it (a)} Quantum or string-loop effects, which are
for sure important for small black holes of mass
 $M\simeq M_{Planck} \simeq
1/\sqrt{\alpha^\prime} g_{string} $, and
{\it (b)} $\sigma$-model or $\alpha^\prime$ corrections, which when
$ g_{string}\ll 1$ can become important much earlier, for
$M\simeq  M_{Planck}/g_{string} \simeq
1/\sqrt{\alpha^\prime} g_{string}^2 $.
Assuming the spectrum stays thermal,
these latter corrections  will  presumably
modify the expression for the temperature $^2$ by replacing
$ G M \rightarrow GM\times f({GM/\sqrt{\alpha^\prime}})$,  where
$ f(y) \simeq 1 + o({1\over y^2})$ for large $y$. Knowledge of the
function $f$, which parametrizes the effects of the
 non-minimal coupling
of strings to gravity,
could be a crucial first step before addressing the paradoxes
related to   black-hole evaporation.

A similar but simpler problem arises when one considers the quantum
instability of strong electromagnetic fields.
Schwinger's field-theoretic calculation $^3$, for minimally-coupled
non-self-interacting particles, gives the following result for the
rate of pair production per unit volume and time:
\begin{equation}
 w =  {2J+1\over 8\pi^3} \sum_{k=1}^{\infty} \ (-)^{(2J+1)(k+1)}\
\Bigl({eE\over k}\Bigr)^2 \exp\Bigl(-{k\pi m^2\over
\vert eE\vert}\Bigr).
\label{schwinger}
\end{equation}
Here  $E$ is the constant external electric field,
  $e$  and $m$ are the charge  and mass of
the produced particles,
and $J = 0$ or ${1\over 2}$ is their spin.
This rate, obtained as the imaginary part of the one-loop
 vacuum-energy density,
vanishes non-perturbatively fast with the external field, showing
that pair-production is a quantum-tunneling phenomenon.
The potential barrier has, roughly-speaking, a height of order $2m$
and a width of order $m/eE$.

This calculation can be also done exactly, if point-particles
 are replaced
by open strings $^4$. The key observation $^5$ is that the
 effect of
a constant electric field $F^{01}= E$ is equivalent to an
 orbifold-like
twist of the light-like coordinates\hfil\break
 $X^{\pm} = {1\over \sqrt{2}} (X^0\pm X^1)$,
by an imaginary angle $\pm i\epsilon$, where
\begin{equation}
 \epsilon = {1\over \pi}  [ arcth(\pi e_1 E) + arcth(\pi e_2 E) ] \ .
\label{twis}
\end{equation}
Here $e_1$ and $e_2$ are   the two charges at the  string endpoints,
and we have set $2{\alpha^\prime} = 1$.
Taking   into account the above twist, and
 the anomalous commutation relation of the
center-of-mass coordinates $ [x^+, x^- ] = {-i \over
 (e_1+e_2)E}$ , one
arrives $^4$  at the following expression for the  annulus
  contribution
to the vacuum energy density of oriented bosonic strings  :
\begin{equation}
  {\cal
F} =  \int_0^\infty {dt\over t}\  (2\pi^2 t)^{-{D\over 2}}\
Z(t)\  f_A(t,e_1 E,e_2 E) \ \ ,
\label{ampli}
\end{equation}
where $ Z(t) = \sum_{states } e^{-{\pi t \over 2}m_S^{\ 2}}$ is the
usual partition function of an open string, $D$ is the
 number of non-compact
space-time dimensions, and the entire field dependence
 is contained in the
correction factor
 \begin{equation}
  f_A = {\pi(e_1+e_2)Et
 e^{-\pi t \epsilon^2 / 2}
\over 2sin(\pi t \epsilon/2)}\
\prod_{n=1}^{\infty}{(1-e^{-\pi t n})^2 \over
  (1-e^{-\pi t(n+i\epsilon)})
(1-e^{-\pi t(n-i\epsilon)}) } \ .
\label{ampli1}
\end{equation}
 This factor goes to one in the zero-field limit,
and has evenly-spaced simple poles at
 $t = {2k\over \vert\epsilon\vert}$
($k=1,2,..$) on the integration axis.
These poles give in turn rise to an imaginary
 part of ${\cal F}$, and hence to
a string pair-production rate
\begin{equation}
w_{bos} =   {1\over 2(2\pi)^{D-1}}\ \sum_{states \ S}
{(e_1+e_2)E\over \epsilon }
\sum_{k=1}^{\infty} (-)^{k+1}
\Bigl({\vert\epsilon\vert\over k}\Bigr)^{D/2}
\exp{\Bigl(-{\pi k \over \vert\epsilon\vert}
( m_S^{\ 2}+\epsilon^2)\Bigr)}\ .
\label{bos}
\end{equation}
The above result stays valid for unoriented strings
 $^4$, for which the
  projection onto physical states is   achieved by adding
to the annulus
the contribution of the M\"obius-strip  $^6$.
 For the fermionic string, on the other hand, a straightforward
extension of the calculation $^4$ gives
\begin{equation}
w_{sustr} =   {1\over 2(2\pi)^{D-1}}\ \sum_{states \ S}
{(e_1+e_2)E\over \epsilon }
\sum_{k=1}^{\infty} (-)^{(k+1)(a_{S}+1)}
\Bigl({\vert\epsilon\vert\over k}\Bigr)^{D/2}
\exp{\Bigl(-{\pi k  m_S^{\ 2}\over \vert\epsilon\vert}  \Bigr)},
\label{ampli1}
\end{equation}
where $a_S=0$ or $1$ according to whether the state
 $S$ belongs to the
Neveu-Schwarz or the Ramond sector.

The string-pair-production rate, eqs.(5,6), reduces
to Schwinger's result,
eq.(1), in the weak-field limit, as has been noticed
before by Burgess
$^7$. The  non-minimal electromagnetic coupling of strings
manifests, however, itself for stronger $E$, and is
  essentially summarized by the
  replacement:
\begin{equation}
 (e_1+e_2)E \rightarrow \epsilon \hfil\break
 \simeq (e_1+e_2)E\times  [1 +
o\Bigl(  ( E \alpha^\prime)^2\Bigr) ]  \ .
\label{repl}
\end{equation}
The non-linear function $\epsilon$, as well as
the string-pair-production rate, diverge  at
 the limiting
field value  $E_{lim} = 1/  2\pi\alpha' max(e_i)   $.
The existence of a limiting field has been noticed also
at the classical level $^{8,5,7 }$, but the above result
clearly establishes the responsible dissipation mechanism.
 Analogous results for strong gravitational, or non-abelian
fields would be extremely interesting. The latter could
in particular drastically modify
 our current semi-phenomenological $^9$
understanding  of fragmentation
in Q.C.D.

Before concluding let me note that
while the string-structure seems to
enhance the effects  of an electric field, it
moderates those of a magnetic field.
 A magnetic field, $F^{12}=B$,
  twists the
  complex space-coordinate $X = {1\over \sqrt{2}}(X^1+iX^2)$
by a real phase $^5$
\begin{equation}
\tilde \epsilon = {1\over \pi}
 [ arctan(\pi e_1 B) + arctan(\pi e_2B) ] \ ,
\label{twist}
\end{equation}
which determines the spacing
 between  Landau levels.
For  sufficiently massive string
states this spacing
reads $\tilde\epsilon/2m$. It   reduces to
the usual cyclotron frequency for weak $B$,
 but levels off towards a
 limiting value
$1/4\alpha^\prime m$ as the field becomes stronger.
This behaviour can be understood as due to a maximum
 allowed centripetal
acceleration.
Notice that the degeneracy of any given Landau level is the same
as for a point particle,
since the commutation relation of the
 center-of-mass coordinates is
not modified $^5$. Finally let me point out that the
  annulus amplitude in the presence of a $B$-field
has no poles on the integration axis . Thus,   as
in point-particle
field theory, a magnetic field cannot dissipate
 by spin-polarizing
the vacuum.
\vglue 0.6cm
This research was  supported in part  by
 EEC contracts SC1-0394-C
and SC1-CT92-0792.

\vglue 0.6cm
{\bf \noindent References \hfil}
\vglue 0.4cm

\end{document}